\documentstyle[]{ptptex}
\def\Journal#1#2#3#4{{#1}{\bf #2} (#4), #3}
\def\PTP{Prog.~Theor.~Phys.~}

\def\NCA{Nuovo~Cimento~}

\def\NPA{Nucl.~Phys. \bf{A}}
\def\NPB{Nucl.~Phys. \bf{B}}
\def\PLB{Phys.~Lett. \bf{B}}

\def\PRL{Phys.~Rev.~Lett.~}
\def\PRD{Phys.~Rev. \bf{D}}
\def\ZPC{Z.~Phys. \bf{C}}

\def\PR{Phys.~Rev.~}
\def\AP{Ann.~Phys.~}
\title{%
Reply to the criticism raised by Pennington\\
on our $\pi\pi$-Production Amplitudes
}
\author{%
Muneyuki {\sc Ishida}, Shin {\sc Ishida}$^A$ and Taku {\sc Ishida}$^B$ 
}
\inst{%
Department of Physics, University of Tokyo , Tokyo 113\\
$^A$ Atomic Energy Research Institute, College of Science and Technology\\
Nihon University, Tokyo 101\\ 
$^B$ KEK, Oho, Tsukuba, Ibaraki 305, JAPAN
}
\recdate{%
\today
}
\abst{%
A recent criticism raised by M.R.Pennington on our 
 $\pi\pi$-production amplitudes seems to be 
due to his misunderstanding on our description, and is not right.
Essential point of the relation between our scattering and production
amplitudes is, in connection to the unitarity and the 
final state interaction theorem, briefly explained.
}

\begin{document}
\maketitle

\setcounter{tocdepth}{4}

Recently, we had analyzed the $\pi\pi$-scattering phase 
shift\cite{rf:pipip,rf:further}\footnote{
Other recent works\cite{rf:achasov,rf:kamin,rf:torn,rf:hara}
 also suggest the existence of the light $\sigma$-particle.} 
and the $\pi\pi$-production
processes, $pp$-central 
collision\cite{rf:taku1,rf:taku2,rf:alde} 
and $J/\Psi\rightarrow\omega\pi\pi$-decay,\cite{rf:JPsi} 
and shown the evidence for existence of 
the long-sought $\sigma$-particle\cite{rf:sca,rf:kunih0}.
These results were reported in 
detail\cite{rf:had97,rf:had97a,rf:had97b,rf:had97d,rf:had97c} 
in the international conference 
HADRON'97 at Brookhaven National Laboratory in August 1997.
However, our description of $\pi\pi$-production processes was
criticized by M. R. Pennington
in Ref. \citen{rf:pen},``The production model of Ishida et al. 
and unitarity": 
It was argued that in our description any production
amplitude has a spurious zero at the same position of 
energy as that in the scattering amplitude.
Here we explain briefly that his criticism 
is unfortunately based on his misunderstanding on our description
 and is not right. 

There are two general problems in treating scattering and production 
amplitudes: the $\pi\pi$-scattering amplitude 
 ${\cal T}$ must satisfy the unitarity relation, and 
the $\pi\pi$-production amplitude ${\cal F}$
must satisfy, in cases with no initial phases,
 the final state interaction(FSI) theorem, 
that is,  ${\cal F}$ must have the same phase\cite{rf:watson} 
as that of ${\cal T}$
.
In order to obtain ${\cal T}$ and ${\cal F}$ satisfying the unitarity and 
the FSI condition, respectively, we start from a 
simple field-theoretical 
model\cite{rf:aitchson,rf:chung,rf:rosen,rf:had97,rf:had97c}:
In the NJL-type model\cite{rf:NJL} as a low energy effective theory 
of QCD, (and in the linear $\sigma$ model, L$\sigma$M,\cite{rf:LsM} 
obtained as its local limit), or in the constituent quark model, 
the pion $\pi$ and the resonant particles such as $\sigma$ or $f_0(980)$
are the color-singlet $q\bar q$-bound states and  
are treated on an equal footing. 
These  ``intrinsic quark dynamics states," denoted as 
$\bar \pi,\ \bar \sigma ,\  \bar f$,
are stable particles with zero widths and appear from the beginning. 
Actually these particles have structures and interact with one another 
and with a production channel ``$P$'' 
through the residual strong interaction:
\begin{eqnarray}
{\cal L}^{\rm scatt}_{\rm int} &=& \bar g_{\bar\sigma}\bar\sigma\pi\pi
+\bar g_{\bar f}\bar f\pi\pi+\bar g_{2\pi}(\pi\pi )^2 \nonumber\\
{\cal L}^{\rm prod}_{\rm int} &=& \bar \xi_{\bar\sigma}\bar\sigma ``P"
+\bar \xi_{\bar f}\bar f``P"+\bar\xi_{2\pi}(\pi\pi ) ``P".
\label{eq:Lint}
\end{eqnarray}
As a result, these bare states change\cite{rf:achasov,rf:torn} into
the physical states, denoted as
$\pi (=\bar\pi ),\ \sigma$ and $f$, with finite widths. 

For simplicity we consider the resonance-dominative 
case, where the background direct two pion coupling 
is neglected( $\bar{g}_{2\pi}=\bar\xi_{2\pi}=0$).
By taking into account only the repetition of pion-loop, 
the ${\cal T}$-matrix automatically satisfies the unitarity.
The relevant two bare states, $\bar\sigma$ and $\bar f$, 
transmit to each other through the 
pion loop diagram, and accordingly the squared mass matrix takes 
a non-diagonal form both in the real and imaginary parts.
By diagonalizing the real part,  ${\cal T}$ is expressed in terms of 
the renormalized masses and coupling constants of the relevant resonances.
Then by a simple manipulation,
 ${\cal T}$ becomes of the same 
form as the conventional ${\cal K}$-matrix representation.
\begin{eqnarray}
{\cal T} 
  &=& 
{\cal K}/(1-i\rho{\cal K});\ \ \ \ 
{\cal K}
=\bar g_{\bar\sigma}(\bar m_{\bar\sigma}^2-s)^{-1}\bar g_{\bar\sigma}
+\bar g_{\bar f}(\bar m_{\bar f}^2-s)^{-1}\bar g_{\bar f}.
\label{eq:Kresrep}
\end{eqnarray}
This ``${\cal K}$-matrix" corresponds to 
the stable particle propagator with an infinitesimal 
imaginary width, and is discriminated from the conventional 
real ${\cal K}$-matrix in the potential theory.
The denominator of Eq.(\ref{eq:Kresrep}) represents the effect of the
repetition of the pion loop. 

The production amplitude  ${\cal F}$, satisfying the FSI-condition, 
is obtained simply by replacing the factor   
 $(\bar{g}_{\bar\alpha})^2\ \ 
({\rm where}\ \bar\alpha =\bar\sigma ,\bar{f})$ 
in ${\cal K}$, appearing  in the numerator of ${\cal T}$,
by $\bar g_{\bar\alpha}\bar\xi_{\bar\alpha}$.
\begin{eqnarray}
{\cal F} &=& {\cal P}/(1-i\rho{\cal K});\ \ \ \ 
{\cal P}=\bar\xi_{\bar\sigma}
(\bar m_{\bar\sigma}^2-s)^{-1}\bar g_{\bar\sigma}
+\bar\xi_{\bar f}
(\bar m_{\bar f}^2-s)^{-1}\bar g_{\bar f},
\label{eq:Presrep}
\end{eqnarray}
where ${\cal P}$ is called the production ``${\cal K}$-matrix."

${\cal T}$ and ${\cal F}$ can be represented directly in terms of 
the physical state bases, and 
${\cal F}$ in the VMW-method\cite{rf:sawa} is reproduced: 
\begin{eqnarray}
{\cal F} &=& 
\frac{r_\sigma e^{i\theta_\sigma}}{\lambda_\sigma-s}  
+\frac{r_f e^{i\theta_f}}{\lambda_f-s},
\label{eq:VMW}
\end{eqnarray}
where $\lambda_\sigma$ and $\lambda_f$ are the (complex) squared masses
of the physical $\sigma$ and $f_0(980)$ particles, respectively. 
Thus, ${\cal F}$ is represented by the sum of Breit-Wigner amplitudes 
of the relevant resonances with 
respective production couplings and phases, $r_\sigma$, $r_f$;  
 $\theta_\sigma$, $\theta_f$.
These new quantities 
are represented by the bare quantities, $\bar{m}_{\bar\alpha}$, 
  $\bar{g}_{\bar\alpha}$, and  $\bar{\xi}_{\bar\alpha}$ 
($\alpha =\sigma ,\ f$).\footnote{
Eq.(\ref{eq:VMW}) includes essentially three new parameters
independent of the scattering process: 
 $r_\sigma$, $r_f$, and relative phase $\theta_f-\theta_\sigma$. 
In the present field theoretical model these are represented  by 
two real production coupling constants, $\bar{\xi}_{\bar\sigma}$ 
and $\bar{\xi}_{\bar f}$. 
Thus, the relative phase parameter is constrained by FSI-theorem.  
However, all the processes induced by strong interaction
generally include unknown strong phases, and correspondingly 
in the actual analyses we are forced to treat this relative phase 
parameter as being free\cite{rf:had97}.
}
Above discussions can be generalized to in the case 
that background couplings take 
non-zero values($\bar{g}_{2\pi}\neq 0$, $\bar{\xi}_{2\pi}\neq 0$)\cite{rf:had97c}, and it is shown that in our relevant case 
the VMW-method is also reproduced, as will be mentioned later.

The zeros in  ${\cal T}$ and ${\cal F}$,
which are determined by the conditions, ${\cal K}=0$ and
${\cal P}=0$, respectively, occur at the respective positions
of squared energy 
\begin{eqnarray}
s=s_0^{\cal T}=\frac{\bar{g}_{\bar\sigma}^2\bar{m}_{\bar{f}}^2+
\bar{g}_{\bar f}^2\bar{m}_{\bar\sigma}^2}
{\bar{g}_{\bar\sigma}^2+\bar{g}_{\bar f}^2} \ \ {\rm and}\ \ 
s=s_0^{\cal F}=
\frac{\bar{g}_{\bar\sigma}\bar{\xi}_{\bar\sigma}\bar{m}_{\bar{f}}^2+
\bar{g}_{\bar f}\bar{\xi}_{\bar f}\bar{m}_{\bar\sigma}^2}
{\bar{g}_{\bar\sigma}\bar{\xi}_{\bar\sigma}
+\bar{g}_{\bar f}\bar{\xi}_{\bar f}}.
\label{eq:zero}
\end{eqnarray}
$s_0^{\cal F}$ is dependent on the production couplings,  
$\bar{\xi}_{\bar\sigma}$ and $\bar{\xi}_{\bar\sigma}$, 
and generally different from $s_0^{\cal T}$(,
and in the case with $\bar{g}_{\bar\sigma}\bar{\xi}_{\bar\sigma}
+\bar{g}_{\bar f}\bar{\xi}_{\bar f}=0$ 
the zero in ${\cal F}$ is removed).
Thus, the criticism, ``a spurious zero (in ${\cal T}$)
transmits to the production processes, 
unphysically shackling its description"\cite{rf:pen},
is clearly incorrect.

That problem had occurred, not in our description,
but in his original scheme of the 
universality\cite{rf:morg,rf:pennington} of $\pi\pi$-scattering 
amplitude ${\cal T}$, 
where the amplitude ${\cal F}$ in any production process is 
represented by (with a smooth real function $\tilde\alpha (s)$ 
of $s$)
\begin{eqnarray}
{\cal F}=\tilde\alpha (s){\cal T}.
\label{eq:FaT}
\end{eqnarray}
Due to this equation, the zeros in  ${\cal T}$ always transmit 
to  ${\cal F}$ at the same positions. 
This is apparently inconsistent with many production experiments.
To avoid this problem, ${\cal T}$ (and correspondingly ${\cal K}$) 
was revised and cleverly replaced by the   
reduced  ${\cal T}$-(${\cal K}$-) matrix, 
$\hat{{\cal T}}(\hat{{\cal K}})$, which is defined by 
\begin{eqnarray}
\hat{{\cal T}}=\frac{{\cal T}}{s-s_0^{\cal T}}\ \ \ 
(\hat{{\cal K}}=\frac{{\cal K}}{s-s_0^{\cal T}}),
\label{eq:red}
\end{eqnarray}
and Eq.(\ref{eq:FaT}) was replaced by 
\begin{eqnarray}
{\cal F}=\alpha (s)\hat{{\cal T}}.
\label{eq:redFaT}
\end{eqnarray}
However, 
this operation seems to be quite artificial and has much arbitrariness,
since we are free to choose any function with the zero at $s=s_0^{\cal T}$,
instead of $s-s_0^{\cal T}$,
to remove the zero of ${\cal T}$(${\cal K}$).
In our scheme we can set up also Eq.(\ref{eq:FaT}) and 
define $\tilde\alpha (s)$. In our case a mechanism corresponding to 
his ``cure" occurs, in contrast with his case,
automatically as follows:
By using Eqs.(\ref{eq:Kresrep}) and (\ref{eq:Presrep}),
$\tilde\alpha (s)$ is determined as, $\tilde\alpha (s)=(
(\bar{g}_{\bar\sigma}\bar{\xi}_{\bar\sigma}
+\bar{g}_{\bar f}\bar{\xi}_{\bar f})
(s-s_0^{\cal F})
)/(
(\bar{g}_{\bar\sigma}^2+\bar{g}_{\bar f}^2)
(s-s_0^{\cal T}) )$, which includes the 
factor $1/(s-s_0^{\cal T})$.

From Eq.(\ref{eq:zero}) we see that $s_0^{\cal F}$ coincides 
with $s_0^{\cal T}$ only in the case,
when there exists a special constraint 
 $\bar{\xi}_{\bar\sigma}/\bar{g}_{\bar\sigma}=
\bar{\xi}_{\bar f}/\bar{g}_{\bar f}$ 
between the production and the scattering couplings, 
which are mutually independent, in principle.
 
Finally we should like to compare concretely the methods of analyses 
between ours and that based on the universality-argument.
In many experiments\cite{rf:taku1,rf:JPsi,rf:CB} 
leading to the $\pi\pi$-channel 
 the spectra of $|{\cal F}|^2$ show a large peak structure  
around the energy $\sqrt{s}\simeq 500$MeV, 
which is quite different from that of $|{\cal T}|^2$
(which has peak at $\sqrt{s}\simeq 900$MeV).
This fact is understood in our VMW method as a result that 
 the background effects are comparatively weaker in 
the production processes than in the scattering process, that is, 
$\bar\xi_{\bar\sigma}/\bar g_{\bar\sigma}  \gg 
\bar\xi_{2\pi}/\bar g_{2\pi}$, when  
the low energy peak structure
is directly reflecting the $\sigma$-resonance.

On the other hand, 
in the conventional analyses based on the 
 ``universality of ${\cal T}$"  
the peak structure is fit by an  
arbitrarily chosen polynomial form 
$\alpha (s)=\Sigma_n \alpha_ns^n$,\cite{rf:morg}
with parameters $\alpha_n$, which has no direct physical meaning.
This situation was clearly explained in ref. \citen{rf:had97}.
 
The difference between the two methods seems to reflect their
basic standpoints: In the universality-argument
only the stable (pion) state consists in 
the complete set of meson
states, while the $\bar \sigma$ and $\bar f$, in addition to the pion, 
are necessary as  bases of the complete set in the VMW-method.


\end{document}